\documentclass[twocolumn]{aastex62}

\usepackage{amsmath}

\received{}
\revised{}
\accepted{}
\submitjournal{ApJL}

\shorttitle{Planetesimal Formation in Dust Rings}
\shortauthors{Stammler et al.}

\newcommand{\figurewidth}{\linewidth}

\begin{document}

\title{The DSHARP Rings: Evidence of Ongoing Planetesimal Formation?}

\correspondingauthor{Sebastian M. Stammler}
\email{stammler@usm.lmu.de}

\author[0000-0002-1589-1796]{Sebastian M. Stammler}
\affil{University Observatory, Faculty of Physics,
Ludwig-Maximilians-Universit\"at M\"unchen, Scheinerstra{\ss}e 1,
D-81679 Munich, Germany}

\author[0000-0002-9128-0305]{Joanna Dr{\c{a}}{\.z}kowska}
\affil{University Observatory, Faculty of Physics,
Ludwig-Maximilians-Universit\"at M\"unchen, Scheinerstra{\ss}e 1,
D-81679 Munich, Germany}

\author[0000-0002-1899-8783]{Til Birnstiel}
\affil{University Observatory, Faculty of Physics,
Ludwig-Maximilians-Universit\"at M\"unchen, Scheinerstra{\ss}e 1,
D-81679 Munich, Germany}

\author[0000-0002-8227-5467]{Hubert Klahr}
\affil{Max-Planck-Institut f\"ur Astronomie, K\"onigstuhl 17,
D-69117 Heidelberg, Germany}

\author[0000-0002-7078-5910]{Cornelis P. Dullemond}
\affil{Zentrum f\"ur Astronomie, Heidelberg University,
Albert-Ueberle-Stra{\ss}e 2, D-69120 Heidelberg, Germany}

\author[0000-0003-2253-2270]{Sean M. Andrews}
\affil{Center for Astrophysics \textbar\ Harvard \& Smithsonian,
60 Garden Street, Cambridge, MA 02138, USA}



\begin{abstract}

Recent high-resolution interferometric observations of protoplanetary disks at (sub-)millimeter wavelengths reveal omnipresent substructures, such as rings, spirals, and asymmetries. A detailed investigation of eight rings detected in five disks by the DSHARP survey came to the conclusion that all rings are just marginally optically thick with optical depths between 0.2 and 0.5 at a wavelength of 1.25\,mm. This surprising result could either be coincidental or indicate that the optical depth in all of the rings is regulated by the same process.

We investigated if ongoing planetesimal formation could explain the "fine-tuned" optical depths in the DSHARP rings by removing dust and transforming it into "invisible" planetesimals. We performed a one-dimensional simulation of dust evolution in the second dust ring of the protoplanetary disk around HD~163296, including radial transport of gas and dust, dust growth and fragmentation, and planetesimal formation via gravitational collapse of sufficiently dense pebble concentrations.

We show that planetesimal formation can naturally explain the observed optical depths if streaming instability regulates the midplane dust-to-gas ratio to unity. Furthermore, our simple monodisperse analytical model supports the hypothesis that planetesimal formation in dust rings should universally limit their optical depth to the observed range.

\end{abstract}

\keywords{accretion, accretion disks ---
          instabilities ---
          methods: numerical ---
          planets and satellites: formation ---
          protoplanetary disks ---
          stars: individual (HD~163296)}


\section{Introduction} \label{sec:introduction}

Since the era of high-resolution interferometry, many circumstellar disks are known to show ring-like substructures in the millimeter continuum emission of
the dust, e.g., HL~Tauri \citep{2015ApJ...808L...3A}, TW~Hydrae \citep{2016ApJ...820L..40A, 2016ApJ...829L..35T}, and HD~163296 \citep{2016PhRvL.117y1101I}.

Recently, the DSHARP survey \citep{2018ApJ...869L..41A} observed 20 protoplanetary disks at a wavelength of 1.25\,mm with an angular resolution of $\sim$0\farcs035. Most of the observed disks show substructures such as rings, spirals, and vortices \citep{2018ApJ...869L..42H, 2018ApJ...869L..43H}.

One of the most promising explanations for substructures in protoplanetary disks is the existence of planets carving gaps in the gas surface density. The outer edge of the planet induced gap acts as a pressure bump and halts the inward drift of dust particles \citep{2006MNRAS.373.1619R, 2012A&A...545A..81P}.

\citet{2018ApJ...869L..46D} analyzed a subset of eight rings in five disks of the DSHARP sample in greater detail and found evidence for dust trapping in pressure bumps. Furthermore, their observations of the azimuthally averaged intensity profiles hint to the existence of a particle size distribution, as apposed to a single grain size. Additionally, a background pressure gradient is needed to account for the deviations from Gaussian profiles in the intensity.

\begin{figure}[tbp!]
  \includegraphics[width=\figurewidth]{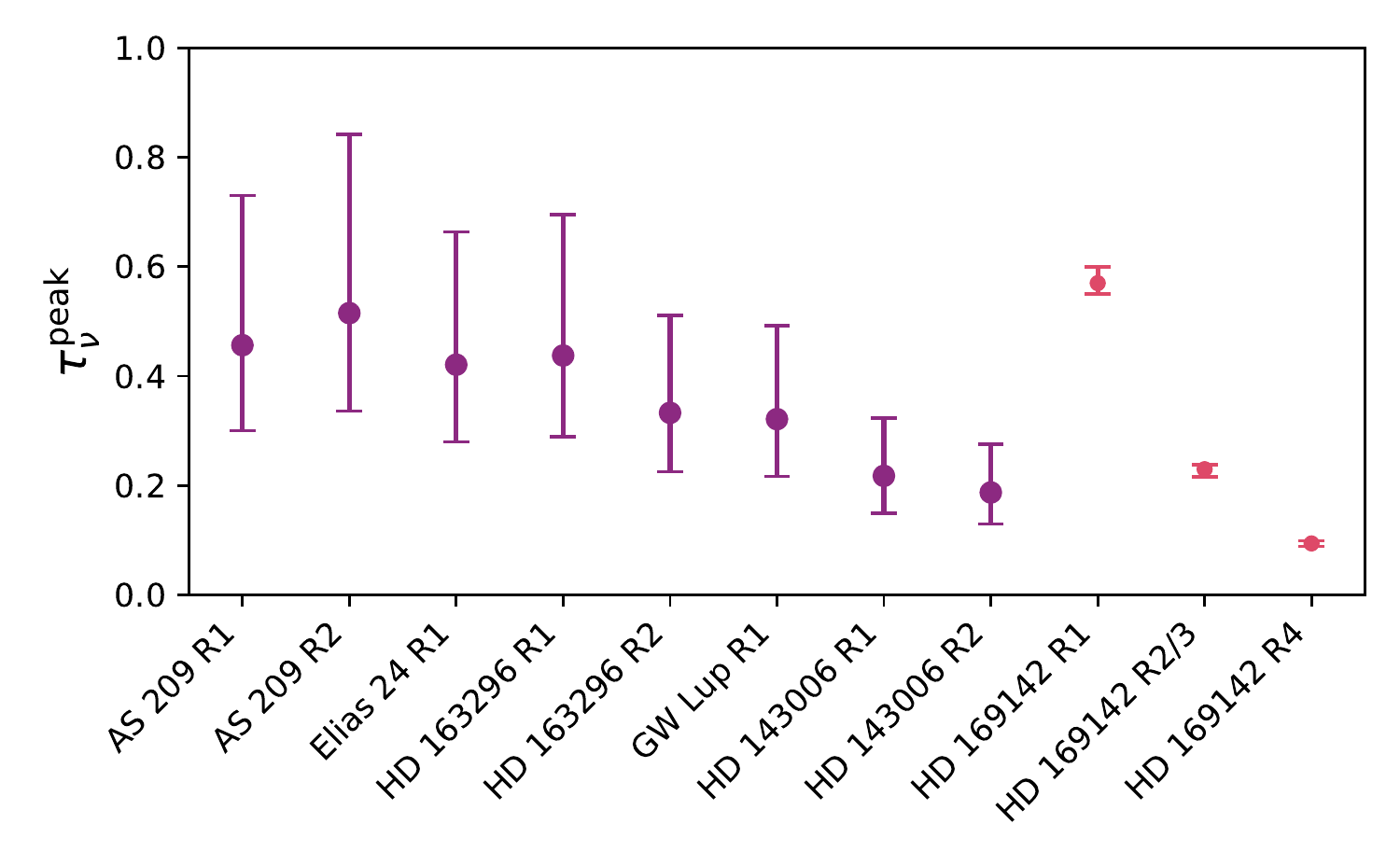}
  \caption{Peak optical depths in the eight rings of the DSHARP sample \citep{2018ApJ...869L..46D} and in the rings of HD~169142 \citep{2019arXiv190707277M}.}
  \label{fig:figure1}
\end{figure}

Another remarkable result of the DSHARP survey is that the derived peak optical depths in the analyzed rings are all very similar: not completely optically thick, but marginally thick with values between about 0.2 and 0.5 (see figure \ref{fig:figure1}). The reason for these seemingly fine-tuned optically thicknesses is unclear. Since \citet{2018ApJ...869L..46D} analyzed only the brightest rings within the DSHARP survey, it makes sense that none of the rings is fully optically thin. It cannot, however, explain why none of the rings is fully optically thick. Since the sample of eight rings is rather small, even coincidence cannot be completely ruled out. Nevertheless, with only a few exceptions the other disks in the DSHARP sample show a similar behavior at the location of substructures \citep{2018ApJ...869L..42H}. Similar results have been obtained by \citet{2018A&A...619A.161C} in the case of HD~135344B and by \citet{2019arXiv190707277M} in HD~169142.

One possible explanation is dust removal by planetesimal formation. Streaming  instability is a hydrodynamical mechanism driven by the relative flow of dust and gas that concentrates dust particles until they collapse under their own gravity, forming 100-kilometer-sized planetesimals \citep{2005ApJ...620..459Y, 2007Natur.448.1022J}. \citet{2009ApJ...704L..75J} found that the planetesimal formation via the streaming instability is conditioned by the vertically integrated dust-to-gas ratio, with a threshold of 0.02 in the case of grains with Stokes numbers between 0.1 and 0.4. \citet{2010ApJ...722.1437B} confirmed this threshold and noticed that only the pebbles of $\mathrm{St} \geq 10^{-2}$ actively clump and thus only the large grains should be taken into account when calculating metallicity to compare with the threshold. \citet{2015A&A...579A..43C} and \citet{2017A&A...606A..80Y} performed systematic studies for the conditions necessary for planetesimal formation and proposed a threshold metallicity criterion as a function of the grain's Stokes number. However, these studies adapted initially laminar disks, where there is only self-driven turbulence. Global disk turbulence can potentially undermine the efficiency of the streaming instability \citep{2018MNRAS.473..796A, 2018ApJ...868...27Y}, however in disks which are turbulent due to a magneto-hydrodynamic instability, zonal flows have been shown to form, which create pressure bumps that concentrate pebbles sufficiently to allow for a spontaneous gravitational collapse \citep{2013ApJ...763..117D}. Also, lower pressure gradients present in pressure bumps have been shown to favour planetesimal formation and the planetesimal formation criterion should depend on the pressure gradient \citep{2010ApJ...722L.220B, 2018arXiv181010018A, 2018ApJ...860..140S}.

While the detailed criteria of conditions allowing for planetesimal formation in the streaming instability in turbulent disks are subject to ongoing studies, in this paper we follow \citet{2016A&A...594A.105D} and \citet{2017A&A...602A..21S} and adopt a simple criterion based on the midplane dust-to-gas ratio exceeding unity, as this seems to be a general criterion for fast growth in the linear phase and development of strong clumping in the non-linear phase of the streaming instability \citep{2007ApJ...662..613Y, 2007ApJ...662..627J}.

The transformation of dust particles into planetesimals could naturally explain the limitation in optical thickness that is observed in dust rings. The self-regulating nature of this process -- a high concentration of dust particles is required and streaming instability might be stalled as soon as enough dust is converted into planetesimals -- could explain why the optical depths in these rings seem to be all within a narrow range.

To investigate this hypothesis, we reproduced the model presented in \citet{2018ApJ...869L..46D} by imposing a Gaussian gap onto the gas which is imitating the gap caused by a planet, but including the full grain size distribution regulated by particle growth and fragmentation. What is more, we implemented a simple recipe for the formation of planetesimals in dust concentrations and analyzed the evolution of the peak optical depth in the dust ring that forms at the outer edge of the gap. Furthermore, we compare this model to a model without planetesimal formation.

In section \ref{sec:analytic}, we derive a simple analytical formula for the optical depth resulting from a monodisperse particle size distribution that is just at the threshold, where streaming instability can act. Section \ref{sec:model} describes the numerical model with which we simulate the growth and transport of dust in protoplanetary disks. In section \ref{sec:results}, we present the results, which are discussed in section \ref{sec:discussion}. In section \ref{sec:alternatives}, we briefly discuss alternative explanations for the observed peak optical depths. We summarize our findings in Section~\ref{sec:last}.

\section{Analytic derivation} \label{sec:analytic}

For a monodisperse dust size distribution, the maximum optical depth can be calculated analytically. The optical depth is the product of the opacity $\kappa_\nu$ and the dust surface density~$\Sigma_\mathrm{d}$
\begin{equation}
  \tau_\nu = \kappa_\nu \Sigma_\mathrm{d}.
  \label{eqn:taunu}
\end{equation}
The opacity $\kappa_\nu$ can be expressed in terms of the dimensionless absorption coefficient $Q_\nu$
\begin{equation}
  \kappa_\nu = \frac{\pi a^2}{m} Q_\nu = \frac{3}{4} \frac{Q_\nu}{a\rho_\mathrm{s}},
\end{equation}
with the particle bulk density $\rho_\mathrm{s}$. The particle size $a$ can be expressed via the dimensionless Stokes number
\begin{equation}
  \mathrm{St} = \frac{\pi}{2} \frac{a\rho_\mathrm{s}}{\Sigma_\mathrm{g}},
\end{equation}
with the gas surface density $\Sigma_\mathrm{g}$.

An important criterion for the streaming instability is the midplane dust-to-gas ratio $\rho_\mathrm{d}/\rho_\mathrm{g}$ \citep{2005ApJ...620..459Y}. We therefore convert the surface
densities to midplane volume densities via
\begin{eqnarray}
  \Sigma_\mathrm{g} &= \sqrt{2\pi} H \rho_\mathrm{g}, \\
  \Sigma_\mathrm{d} &= \sqrt{2\pi} h \rho_\mathrm{d},
\end{eqnarray}
with the pressure scale height of the gas $H$ and the dust scale height $h$, which is given by \citet{1995Icar..114..237D} as
\begin{equation}
  h = \sqrt{\frac{\alpha}{\alpha + \mathrm{St}}} H
  \simeq \sqrt{\frac{\alpha}{\mathrm{St}}} H,
\end{equation}
where $\alpha$ is the viscosity parameter \citep{1973A&A....24..337S}. The last step is an approximation for $\alpha \ll \mathrm{St}$.

Putting everything into equation (\ref{eqn:taunu}) results in
\begin{equation}
  \tau_\nu = \frac{3\pi}{8} \sqrt{\frac{\alpha}{\mathrm{St}^3}} Q_\nu \frac{\rho_\mathrm{d}}{\rho_\mathrm{g}}.
\end{equation}
For the threshold midplane dust-to-gas ratio of unity and reasonable values of the needed quantities, the resulting equation reads
\begin{equation}
  \tau_\nu = 0.5\ \frac{Q_\nu}{0.4}\ \left( \frac{\alpha}{0.001} \right)^\frac{1}{2}\ \left( \frac{\mathrm{St}}{0.1} \right)^{-\frac{3}{2}}.
  \label{eqn:peakTauAnalyic}
\end{equation}
The value of $Q_\nu$ was calculated by using the DSHARP opacities \citep{2018ApJ...869L..45B} and the local conditions in the dust ring in the simulation presented in section \ref{sec:results}. While equation (\ref{eqn:peakTauAnalyic}) has a rather steep dependence on the Stokes number, we would like to point out, that $Q_\nu$ itself depends on the particle size and therefore the Stokes number. To first order approximation $Q_\nu \propto a \propto \mathrm{St}$ in a regime where $\lambda > 2 \pi a$ \citep{1997MNRAS.291..121I}. This lowers the effective dependence on the Stokes number.

When planetesimal formation is able to keep the midplane dust-to-gas ratio at unity, this could be a natural explanation for the marginally optically thick dust rings in the DSHARP survey. The idea is that, as soon as the dust surface density (and thereby the optical depth) exceeds this threshold, particle concentration sets in \citep{2018ApJ...861...47S}. Clumps of dust form, which gravitationally collapse to form planetesimals. This takes mass away from the dust population, lowering the dust surface density, and shutting down the streaming instability again. This self-regulated process will thus keep the dust surface density right at the border of stability, and thus keep the optical depth close to the value given by equation~(\ref{eqn:peakTauAnalyic}).

However, this simple expression is only valid for a single particle size. For a more detailed analysis with a particle size distribution, we performed full numerical models.

\section{Numerical Model} \label{sec:model}

We modeled the second dust ring of HD~163296 in a similar way as \citet{2018ApJ...869L..46D}. The one-dimensional simulations have been performed with \texttt{DustPy}, a \texttt{Python}-based software package for dust growth and evolution in protoplanetary disks, which is based on the model of \citet{2010A&A...513A..79B}. We imposed a Gaussian-shaped gap onto the gas by increasing the viscosity in this region respectively. Gas and dust dynamics have been implemented by solving their continuity equations. We followed grain growth and fragmentation by solving the Smoluchowski equation with a simple sticking-fragmentation collision model. To account for planetesimal formation by streaming instability, we removed mass from the dust distribution with a simple recipe.

All input parameters of our model are listed in table~\ref{tab:inputs}. Note that the radial mixing parameter $\delta$ is a factor of two larger than the viscosity parameter $\alpha_0$ to reproduce the observed width of the dust ring.

\begin{deluxetable}{llrl}[tbp!]
\tablecaption{Input parameters of the model. \label{tab:inputs}}
\tablecolumns{4}
\tablehead{
\colhead{Symbol} &
\colhead{Description} &
\colhead{Value} &
\colhead{Unit}
}
\startdata
$\alpha_0$ & viscosity parameter & $0.001$ & -- \\
$\delta$ & radial mixing parameter & $0.002$ & -- \\
$\epsilon$ & efficiency of planetesimal formation & $0.1$ & -- \\
$f$ & gap depth & 2.0 & -- \\
$\gamma$ & slope of surface density & 1.0 & -- \\
$L_*$ & stellar luminosity & 17.0 & $L_\sun$ \\
$M_*$ & stellar mass & 2.04 & $M_\sun$ \\
$M_\mathrm{disk}$ & initial disk mass & 0.4 & $M_\sun$ \\
$\varphi$ & irradiation angle & 0.02 & rad \\
$r_\mathrm{c}$ & critical cut-off radius & 200 & AU \\
$r_\mathrm{p}$ & gap position & 83.5 & AU \\
$\rho_\mathrm{s}$ & particle bulk density & 1.6 & g/cm$^3$ \\
$\Sigma_\mathrm{d}/\Sigma_\mathrm{g}$ & dust-to-gas ratio & 0.01 & -- \\
$v_\mathrm{f}$ & fragmentation velocity & 10.0 & m/s \\
$w_\mathrm{gap}$ & gap width & 6 & AU
\enddata
\end{deluxetable}

\subsection{Gas and dust dynamics}

We initially set the gas disk according the self-similar solution of \citet{1974MNRAS.168..603L}:

\begin{equation}
  \Sigma_\mathrm{g} \left( r \right) = \Sigma_0 \left( \frac{r}{r_\mathrm{c}} \right)^{-\gamma} \exp \left[ \left( -\frac{r}{r_\mathrm{c}} \right)^{2-\gamma} \right].
\end{equation}
The parameter $\Sigma_0 = \left( 2 - \gamma \right) M_\mathrm{disk} / \left( 2 \pi r_\mathrm{c}^2 \right)$ is set by the initial disk mass.

The initial dust distribution follows the gas distribution with a constant dust-to-gas ratio. The initial particles sizes follow the distribution of the interstellar medium \citep{1977ApJ...217..425M} with a maximum particle size of 1\,$\mu$m.

We follow the gas evolution by solving the continuity equation
\begin{equation}
  \frac{\partial}{\partial t} \Sigma_\mathrm{g} + \frac{1}{r} \frac{\partial}{\partial r} \left( r \Sigma_\mathrm{g} v_{\mathrm{g},r} \right) = 0,
\end{equation}
where the radial gas velocity is given by
\begin{equation}
  v_{\mathrm{g},r} = - \frac{3}{\Sigma_\mathrm{g}\sqrt{r}} \frac{\partial}{\partial r} \left( \Sigma_\mathrm{g}\sqrt{r} \nu \right),
\end{equation}
with $\nu = \alpha c_\mathrm{s}^2 / \Omega_\mathrm{K}$ being the turbulent viscosity, $\alpha$ the viscosity parameter, $c_\mathrm{s}$ the sound speed, and $\Omega_\mathrm{K}$ the Keplerian frequency.

Every dust particle size $i$ follows its own advection-diffusion equation:
\begin{equation}
  \frac{\partial}{\partial t} \Sigma_\mathrm{d}^i + \frac{1}{r} \frac{\partial}{\partial r} \left( r \Sigma_\mathrm{d}^i v_{\mathrm{d},r}^i \right)
  = \frac{1}{r}\frac{\partial}{\partial r} \left[ rD^i\Sigma_\mathrm{g} \frac{\partial}{\partial r} \left( \frac{\Sigma_\mathrm{d}^i}{\Sigma_\mathrm{g}} \right) \right],
\end{equation}
where the dust diffusivity is given by \citet{2007Icar..192..588Y} as
\begin{equation}
  D^i = \frac{\delta c_\mathrm{s}^2 / \Omega_\mathrm{K}}{1+{\mathrm{St}^i}^2}.
\end{equation}
$\delta$ is the radial mixing parameter, similar to $\alpha$ for the gas evolution. The radial dust velocity is
\begin{equation}
  v_{\mathrm{d},r}^i = \frac{1}{1+{\mathrm{St}^i}^2} v_{\mathrm{g},r} + \frac{1}{{\mathrm{St}^i}+1/{\mathrm{St}^i}} \frac{c_\mathrm{s}^2}{\Omega_\mathrm{K}r}
  \frac{\mathrm{d}\ln p}{\mathrm{d}\ln r},
\end{equation}
where $p$ is the gas pressure. The Stokes number is defined as
\begin{equation}
  \label{eq:stokesNumber}
  \mathrm{St}^i = \frac{\pi}{2} \frac{a^i\rho_\mathrm{s}}{\Sigma_\mathrm{g}},
\end{equation}
with the particle radii $a^i$ and the particle bulk density $\rho_\mathrm{s}$.

\subsection{Dust growth}

We simulate dust growth by following the particle mass distribution $f\left( m \right)$. This is done by solving the Smoluchowski equation
\begin{equation}
  \frac{\partial}{\partial t} f \left( m \right)
  = \iint f \left( m^\prime \right) f \left( m^{\prime\prime} \right) M \left(m, m^\prime, m^{\prime\prime} \right) \mathrm{d}m^{\prime\prime} \mathrm{d}m^\prime,
\end{equation}
with the coagulation Kernel $M \left(m, m^\prime, m^{\prime\prime} \right)$. Particles grow by hit-and-stick collisions until their relative collision velocities exceed the fragmentation velocity $v_\mathrm{f}$, where they start to fragment. The exact collisional physics are hidden in the coagulation Kernel. For a detailed description of the coagulation/fragmentation method used here we refer to \citet{2010A&A...513A..79B}.

\subsection{Temperature profile}

For the temperature profile, we assume a simple irradiated disk model with the midplane temperature given by
\begin{equation}
  T \left( r \right) = \left( \frac{\frac{1}{2} \varphi L_*}{4 \pi r^2 \sigma_\mathrm{SB}} \right)^{1/4},
\end{equation}
with the stellar luminosity $L_*$, the Stefan-Boltzmann constant $\sigma_\mathrm{SB}$, and the irradiation angle $\varphi$. We assume that gas and dust are always well-coupled and share the same temperature. Further, we assume that the temperature does not change with height above the midplane. The stellar luminosity does not change during our simulation.

\subsection{Streaming instability}

Since we cannot self-consistently solve for the hydrodynamical interactions between dust and gas leading to the streaming instability in our one-dimensional model, we implemented a simple recipe for forming planetesimals in dust concentrations \citep[see, e.g.,][]{2016A&A...594A.105D, 2018A&A...620A.134S}. As soon as the midplane dust-to-gas ratio exceeds unity, we remove a fraction $\epsilon=0.1$ of the dust surface density per settling timescale and shift this mass into the surface density of planetesimals. The rate of change $R^i$ per species is then given by
\begin{equation}
  R^i = \frac{\partial}{\partial t} \Sigma_\mathrm{d}^i = - \epsilon \frac{\Sigma_\mathrm{d}^i}{t_\mathrm{sett}^i}
  = - \epsilon \Sigma_\mathrm{d}^i \mathrm{St}^i \Omega_\mathrm{K}.
\end{equation}
The mass that gets added to the planetesimals is then simply the sum over all dust sizes
\begin{equation}
  \frac{\partial}{\partial t} \Sigma_\mathrm{pl} = -\sum_i R^i.
\end{equation}
We do not further evolve the surface density of planetesimals.

\subsection{Gas gap}

To model a gap carved by a planet we follow the approach of \citet{2018ApJ...869L..46D}. Since in the steady state $\alpha\cdot\Sigma_{g}$ is constant, a method for inducing a gap in the gas density is to have a bump in the $\alpha$ viscosity parameter
\begin{equation}
  \alpha \left( r \right) = \frac{\alpha_0}{F\left( r \right)},
\end{equation}
where the function $F\left( r \right)$ is given by
\begin{equation}
  F \left( r \right) = \exp \left[ -f \exp \left( -\frac{\left( r - r_\mathrm{p} \right)^2}{2w_\mathrm{gap}^2} \right) \right].
\end{equation}
This only changes the turbulent viscosity of the gas. The radial mixing of the dust or the calculation of the turbulent collision velocity of the dust particles is not affected by this modification.

\subsection{Optical properties}

To calculate the optical depth, the intensity profiles, and the spectral index we use the DSHARP opacity model \citep{2018ApJ...869L..45B}, which uses optical constants of water ice from \citet{2008JGRD..11314220W}, of astronomical silicates from \citet{2003ApJ...598.1026D}, and of troilite and organics from \citet{1996A&A...311..291H}.

\section{Results} \label{sec:results}

\begin{figure}[tbp!]
  \includegraphics[width=\figurewidth]{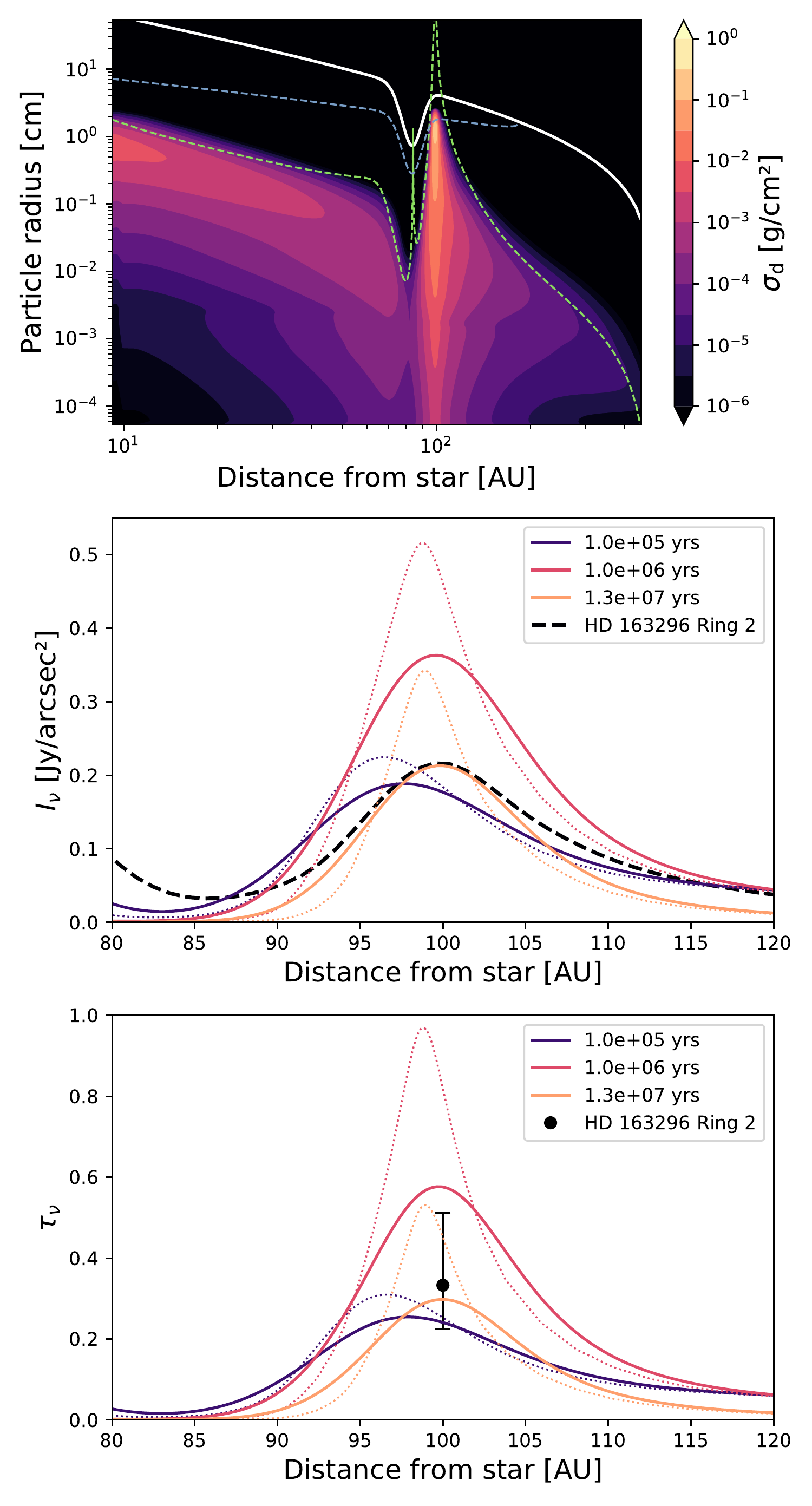}
  \caption{{\it Top panel}: Snapshot of the dust distribution after 13\,Myrs in the simulation with planetesimal formation. The white line shows particles sizes with St=1. The dashed blue and green lines correspond to the fragmentation and drift limits, respectively \citep{2012A&A...539A.148B}. {\it Middle panel}: Solid lines show the modeled intensity profiles at 1.25\,mm convolved with the beam size at different snapshots. The dotted lines show the corresponding unconvolved intensity profiles. The dashed black line shows the observed intensity profile \citep{2018ApJ...869L..41A}. {\it Bottom panel}: Optical depth profiles calculated from the convolved (solid) and unconvolved (dotted) intensity profiles using the DSHARP opacity model \citep{2018ApJ...869L..45B}. The data point corresponds to the derived optical depth in the second dust ring of HD~163296 \citep{2018ApJ...869L..46D}.}
  \label{fig:figure2}
\end{figure}

We performed two simulations evolving dust for several million years each. The first simulation included planetesimal formation through the streaming instability, and the second one is a control case without planetesimal formation. The top panel of figure~\ref{fig:figure2} shows the dust surface density distribution of the simulation with planetesimal formation after 13\,Myrs. The plotted quantity $\sigma_\mathrm{d}$ corresponds to the dust surface density of each logarithmic size bin:
\begin{equation}
  \Sigma_\mathrm{d} \left( r \right) = \int\limits_0^\infty \sigma_\mathrm{d} \left( r, a \right) \mathrm{d} \ln a.
\end{equation}
The white solid line representing $\mathrm{St}=1$ particles is proportional to the gas surface density (cf. equation (\ref{eq:stokesNumber})) and shows the gap carved in the gas by a hypothetical planet at 83.5\,AU. At this stage of the simulation, the particles sizes are limited by radial drift everywhere in the disk except for the dust trap at the outer edge of the gas gap at about 100\,AU, where the particles are limited by fragmentation. Particles in the pressure trap reach maximum sizes of about 3\,cm, which corresponds to a Stokes number of about 0.5. Outside the dust ring the particles are limited to a few millimeters or less in size. The vertically integrated dust-to-gas ratio in the pressure bump is about 6\,\%.

We calculated the intensity profile by solving the radiative transfer equation
\begin{equation}
  I_\nu \left( r \right) = \left( 1 - e^{-\tau_\nu\left( r \right)} \right) B_\nu \left( T\left( r \right) \right),
\end{equation}
with the Planck function $B_\nu$ and the optical depth $\tau_\nu$, which is computed using the DSHARP opacity model \citep{2018ApJ...869L..45B}. The middle panel of figure \ref{fig:figure2} shows the intensity profiles at a wavelength of 1.25\,mm at different snapshots in the region of the dust ring. The intensity profile has been convolved with a Gaussian filter with the size of the beam $\sigma_\mathrm{b} = 3.3375\,\mathrm{AU}$ used in the observations of \citet{2018ApJ...869L..41A}. The unconvolved intensity profiles are plotted with dotted lines. The black dashed line is the observed intensity profile, which should be compared to the convolved profiles. The snapshot at 13\,Myrs fits the observed intensity profile best, while it still lacks emission in the outer wings of the bump.

The bottom panel of figure \ref{fig:figure2} shows the corresponding optical depth profiles at 1.25\,mm at the same snapshots. The optical depths has been calculated from the convolved (solid) and unconvolved (dotted) intensity profiles. The data point corresponds to the peak optical depth in the second ring of HD~163296 and its error derived in \citet{2018ApJ...869L..46D}. Again, the peak optical depth at 13\,Myrs in the simulation fits best to the observation. However, the model lies within the error bars for almost the entire lifetime of the protoplanetary disk, between 2\, Myr up to 20\,Myrs.

\begin{figure}[tbp!]
  \includegraphics[width=\figurewidth]{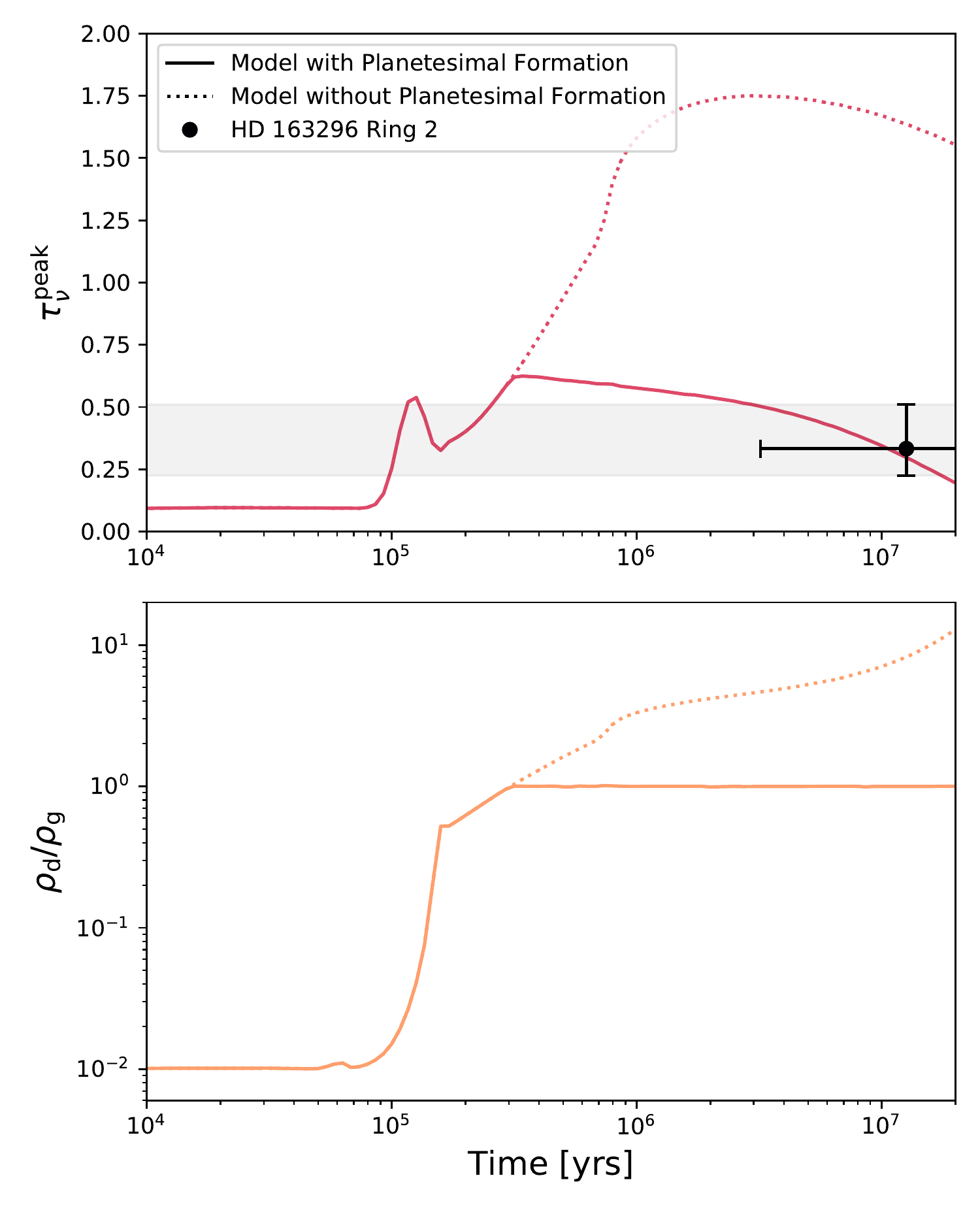}
  \caption{{\it Top panel:} Time evolution of the peak optical depth in the dust ring calculated from the convolved intensity profiles. The data point corresponds to the second dust ring in HD~163296 \citep{2018ApJ...869L..41A, 2018ApJ...869L..46D}. {\it Bottom panel:} Time evolution of the peak midplane dust-to-gas ratio in the dust ring. The solid lines correspond to the model with and the dotted line to the model without planetesimal formation in each panel.}
  \label{fig:peakOpticalDepth}
\end{figure}

Figure \ref{fig:peakOpticalDepth} shows the time evolution of the peak optical depth in the dust ring calculated from the convolved intensity profile and the maximum midplane dust-to-gas ratio. The control simulation without streaming instability is plotted for comparison. In the dust ring, streaming instability sets in after about 300\,000\,yrs, when the dust-to-gas ratio in the midplane reaches the threshold value of unity. At this point the optical depth levels off and stays within the error bars derived from the observations for almost the whole lifetime of the disk. The optical depth in the control case without streaming instability, on the other hand, continues to rise up to values of 1.75. Also the midplane dust-to-gas ratio is stabilized after the streaming instability sets in thanks to its self-regulating nature. In the control case without streaming instability the midplane dust-to-gas ratio reaches values as high as 10.

The peak that is seen in the optical depth in figure \ref{fig:peakOpticalDepth} shortly after 100\,000\,years marks the point in time when the particles hit the fragmentation barrier. Fragmentation limited particles roughly resemble a power law size distribution from the maximum particles size down to monomers. The size distribution of particles that have not yet hit the fragmentation limit is rather comparable to a Gaussian \citep[see e.g.][]{2012A&A...540A..73W}. This influences the resulting opacity of the particle distribution, with the fragmentation limited distribution being slightly less opaque, causing the drop in figure \ref{fig:peakOpticalDepth}. after 150\,000\,years.

\begin{figure}[tbp!]
  \includegraphics[width=\figurewidth]{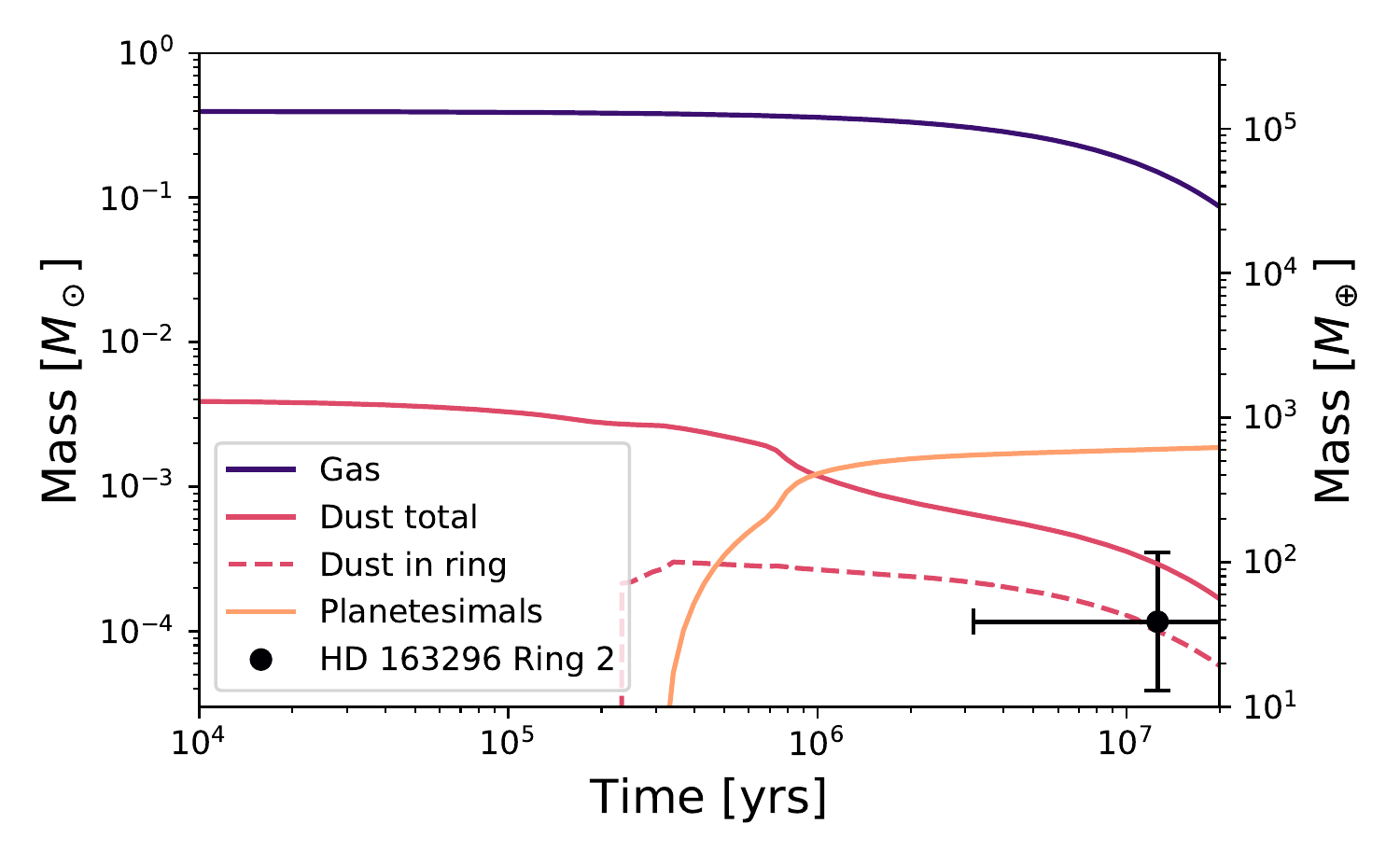}
  \caption{Mass budget of gas, total dust, dust in the ring, and planetesimals. The data point corresponds to the observed dust mass in the ring derived by the DSHARP survey \citep{2018ApJ...869L..41A, 2018ApJ...869L..46D}.}
  \label{fig:massBudget}
\end{figure}

Figure \ref{fig:massBudget} shows the mass budget of gas, dust, and planetesimals during the simulation. The dashed red line represents the dust mass in the ring, where the ring size is defined by the full width at half maximum of the dust surface density. This value should be compared to the data point which corresponds to the dust mass in the ring as estimated by \citet{2018ApJ...869L..46D}.

Planetesimal formation starts after about 300\,000\,yrs and about 600 Earth masses of planetesimals are produced until the end of the simulation. After about 6\,Myrs, 95\,\% of the planetesimals have been formed. The planetesimal were not evolved any further, but simply stayed at the location of their formation.

\begin{figure}[tbp!]
  \includegraphics[width=\figurewidth]{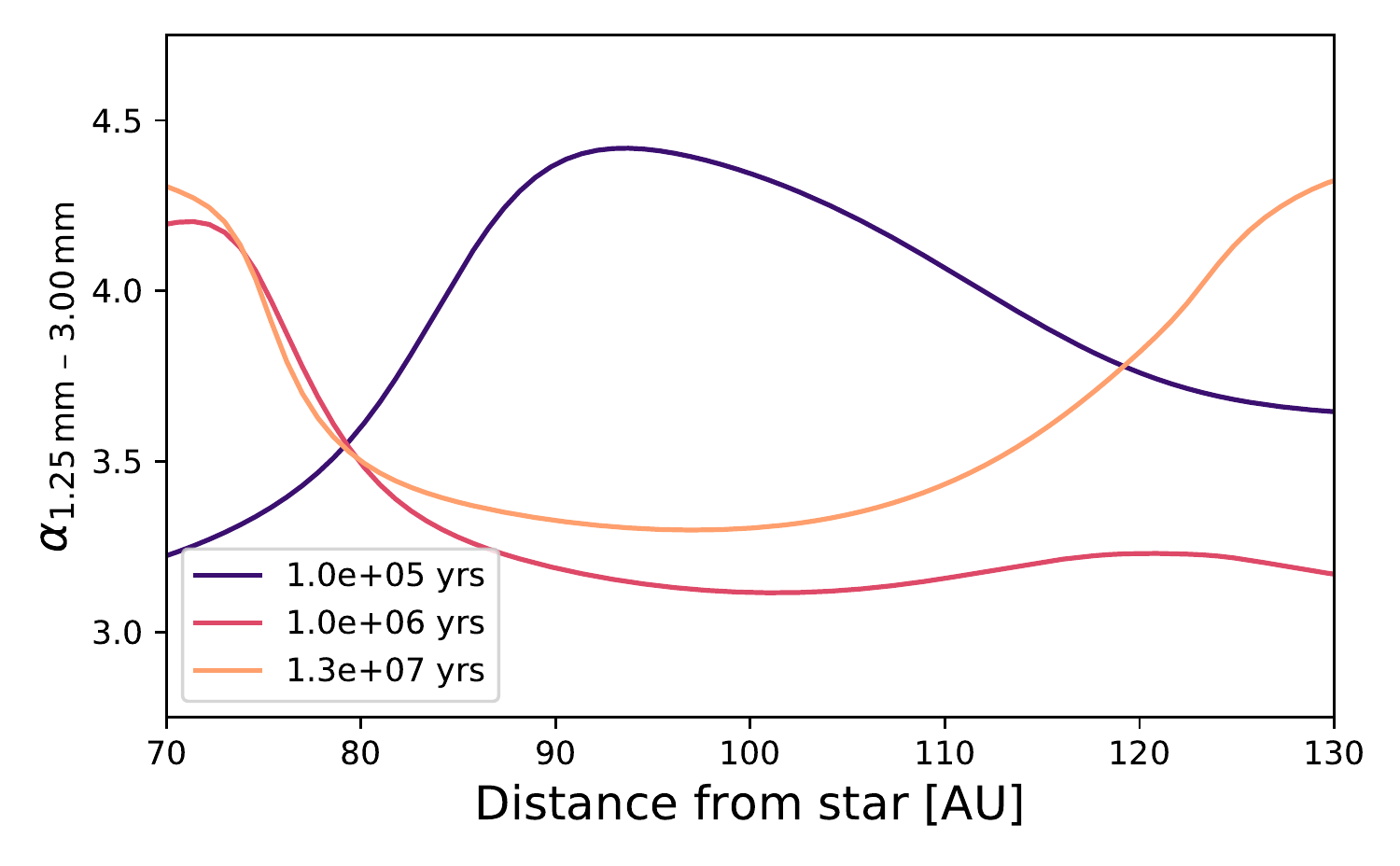}
  \caption{Spectral index profile in the dust ring region at different snapshots calculated from the convolved intensity profiles.}
  \label{fig:spectralIndexProfile}
\end{figure}

Figure \ref{fig:spectralIndexProfile} shows the spectral index in the ring at different snapshots. The spectral index has been calculated using the intensity profiles at the two wavelengths 1.25\,mm and 3.00\,mm convolved with the beamsize at 3\,mm. As soon as fragmentation sets in and the particle size distribution roughly resembles a power law, the spectral index in the dust ring reaches its minimum values between 3.0 and 3.5, while it is significantly higher outside the dust ring later in the simulation.

\section{Discussion} \label{sec:discussion}

\citet{2018ApJ...869L..46D} analyzed eight bright rings observed in the DSHARP survey \citep{2018ApJ...869L..41A}. They found that the rings seen in the continuum emission can be explained by dust particles trapped in pressure bumps. The deviation from a Gaussian intensity profile can be explained by particle size distribution, the asymmetry by a background pressure gradient.

We see the same behavior in our numerical model concerning the outer ring of HD~163296 and including the full size distribution regulated by dust growth and fragmentation. But even in the snapshot at 13\,Myrs, which fits the observations best, the wings of the intensity bump are significantly lower than the observation. However, we only simulated one gap carved by a planet at 83.5\,AU. HD~163296 may have at least three more planets at 50\,AU \citep{2016PhRvL.117y1101I}, at 137\,AU \citep{2018ApJ...860L..12T}, and at 260\,AU \citep{2018ApJ...860L..13P}. If there are additional dust traps inside and outside of the pressure bump simulated in this work, the excess in emission that is observed may be explained by this. Modeling of multiple gaps and a deeper study of disk parameters will be part of future studies.

In our simulation, gravitational collapse of locally concentrated pebbles regulated by streaming instability leads to the formation of more than a Jupiter mass in planetesimals in a narrow ring. We did not further simulate the evolution of these planetesimals, but just let them stay at the location of their formation. Merging, scattering or pebble accretion onto planetesimals was not taken into account and will be a part of future works. Since we did not model any other planet farther outside in the disk, the dust initially located in the outer disk could drift to the dust ring thus increasing the formation rate of planetesimals. A pressure bump in the outer disk could trap some of the dust and lower the drift rate thereby reducing the final mass of planetesimals in the modeled ring.

Following the steps outlined in \citet{2017ASSL..445..197O}, we can estimate the pebble accretion rate onto the planetesimals in the dust ring. Assuming a typical planetesimal size of 100\,km, a single planetesimal would accrete $\sim 10^{-9}\,\mathrm{M}_\oplus/\mathrm{yr}$ using the particle size distribution and densities in the dust ring at 13\,Myrs. At this time we have about $10^{8}$ planetesimals in the simulation, leading to a total pebble accretion rate of $\sim 0.1\,\mathrm{M}_\oplus/\mathrm{yr}$. This is significant compared to the peak planetesimal formation rate of $\sim 10^{-3}\,\mathrm{M}_\oplus/\mathrm{yr}$ in the early simulation and should be taken into account in future works.

\subsection{Alternative explanations}
\label{sec:alternatives}

The formation of planetesimals from small dust particles is not the only possible explanation for the seemingly fine-tuned optical depths in the DSHARP rings. The back reaction of dust particles onto the gas can also smear out concentrations (\citealt{2016A&A...591A..86T, 2019arXiv190607708G}). This effect was not taken into account in this work. Back reaction can usually be neglected, if the dust-to-gas ratio is much lower than unity. But as seen in figure \ref{fig:peakOpticalDepth}, the midplane dust-to-gas ratio in the simulation without planetesimal formation reaches values of about 10. The influence of the back reaction of dust onto the gas on the appearance of the dust rings and the optical depth will be part of future works.

Recent publications have indicated that not only the absorption, but also the scattering opacity plays a significant role in the interpretation of (sub-)millimeter observations \citep{2015ApJ...809...78K, 2019arXiv190400333L, 2019arXiv190402127Z}. Whether the inclusion of scattering effects in the radiative transfer formalism can have a significant influence on the perceived optical depths will be part of future investigations. Observations with longer wavelengths (e.g. with facilities like the ngVLA, \citealp{2018arXiv180304467R}) could help to distinguish the scattering from the planetesimal formation scenario, since scattering effects are highly wavelength-dependent and suppressed at long wavelengths.

\section{Summary}
\label{sec:last}

In this publication, we show that a natural explanation for the peculiar optical depths observed in dust rings in protoplanetary disks is the formation of planetesimals converting small dust into large bodies. A simple analytical derivation assuming a single particle size (see equation~\ref{eqn:peakTauAnalyic}) shows that the optical depth of $\sim0.5$ is naturally obtained if the dust density is regulated by planetesimal formation. This would mean that the observed narrow distribution of optical depths in dust rings can be evidence of ongoing planetesimal formation.

As long as the streaming instability is acting, the midplane dust-to-gas ratio is limited to unity by formation of planetesimals from pebbles. This naturally limits the peak optical depth in the dust ring to values reported by the DSHARP survey for almost the whole lifetime of the protoplanetary disk. Additionally, the dust mass in the ring compares well to the value derived by \citet{2018ApJ...869L..46D}, which is a consequence of using the same opacities with a model that also reproduces the emission.

In a future works we aim to explore a larger parameter space to confirm if planetesimal formation can explain the other rings in the DSHARP survey.

\acknowledgments

We thank the anonymous referee for the helpful comments. S.M.S, J.D., and T.B. have received funding from the European Research Council (ERC) under the European Union's Horizon 2020 research and innovation program under grant agreement No 714769.
S.M.S, J.D., T.B., and C.P.D. acknowledge funding by the Deutsche Forschungsgemeinschaft (DFG, German Research Foundation) ref. no. FOR 2634/1.
S.M.A. acknowledges funding from the National Aeronautics and Space Administration under grant No.~17-XRP17 2-0012 issued through the Exoplanets Research Program.
Part of this work was performed by S.M.S., J.D., and T.B. at the Aspen Center for Physics, which is supported by National Science Foundation grant PHY-1607611.
The authors gratefully acknowledge the compute and data resources provided by the Leibniz Supercomputing Centre (\href{http://www.lrz.de/}{www.lrz.de}).

H.K. received funding by the Deutsche Forschungsgemeinschaft (DFG,  German  Research  Foundation) as part of the Schwerpunktprogramm (SPP, Priority Program) SPP 1833 "Building a Habitable Earth",  Priority Programme "Exploring the Diversity of Extrasolar Planets" (SPP 1992) and in part at KITP Santa Barbara by the National Science Foundation under Grant No. NSF PHY17-48958. Part of this work was performed at the Aspen Center for Physics, which is supported by National Science Foundation grant PHY-1607761. This research was supported by the Munich Institute for Astro- and Particle Physics (MIAPP) of the DFG cluster of excellence "Origin and Structure of the Universe".

\bibliographystyle{aasjournal}
\bibliography{bibliography}

\begin{thebibliography}{}
\expandafter\ifx\csname natexlab\endcsname\relax\def\natexlab#1{#1}\fi
\providecommand{\url}[1]{\href{#1}{#1}}

\bibitem[{{Abod} {et~al.}(2018){Abod}, {Simon}, {Li}, {Armitage}, {Youdin}, \&
  {Kretke}}]{2018arXiv181010018A}
{Abod}, C.~P., {Simon}, J.~B., {Li}, R., {et~al.} 2018, arXiv e-prints,
  arXiv:1810.10018

\bibitem[{{ALMA Partnership} {et~al.}(2015){ALMA Partnership}, {Brogan},
  {P{\'e}rez}, {Hunter}, {Dent}, {Hales}, {Hills}, {Corder}, {Fomalont},
  {Vlahakis}, {Asaki}, {Barkats}, {Hirota}, {Hodge}, {Impellizzeri}, {Kneissl},
  {Liuzzo}, {Lucas}, {Marcelino}, {Matsushita}, {Nakanishi}, {Phillips},
  {Richards}, {Toledo}, {Aladro}, {Broguiere}, {Cortes}, {Cortes}, {Espada},
  {Galarza}, {Garcia-Appadoo}, {Guzman-Ramirez}, {Humphreys}, {Jung}, {Kameno},
  {Laing}, {Leon}, {Marconi}, {Mignano}, {Nikolic}, {Nyman}, {Radiszcz},
  {Remijan}, {Rod{\'o}n}, {Sawada}, {Takahashi}, {Tilanus}, {Vila Vilaro},
  {Watson}, {Wiklind}, {Akiyama}, {Chapillon}, {de Gregorio-Monsalvo}, {Di
  Francesco}, {Gueth}, {Kawamura}, {Lee}, {Nguyen Luong}, {Mangum}, {Pietu},
  {Sanhueza}, {Saigo}, {Takakuwa}, {Ubach}, {van Kempen}, {Wootten},
  {Castro-Carrizo}, {Francke}, {Gallardo}, {Garcia}, {Gonzalez}, {Hill},
  {Kaminski}, {Kurono}, {Liu}, {Lopez}, {Morales}, {Plarre}, {Schieven},
  {Testi}, {Videla}, {Villard}, {Andreani}, {Hibbard}, \&
  {Tatematsu}}]{2015ApJ...808L...3A}
{ALMA Partnership}, {Brogan}, C.~L., {P{\'e}rez}, L.~M., {et~al.} 2015, \apj,
  808, L3

\bibitem[{{Andrews} {et~al.}(2016){Andrews}, {Wilner}, {Zhu}, {Birnstiel},
  {Carpenter}, {P{\'e}rez}, {Bai}, {{\"O}berg}, {Hughes}, {Isella}, \&
  {Ricci}}]{2016ApJ...820L..40A}
{Andrews}, S.~M., {Wilner}, D.~J., {Zhu}, Z., {et~al.} 2016, \apj, 820, L40

\bibitem[{{Andrews} {et~al.}(2018){Andrews}, {Huang}, {P{\'e}rez}, {Isella},
  {Dullemond}, {Kurtovic}, {Guzm{\'a}n}, {Carpenter}, {Wilner}, {Zhang}, {Zhu},
  {Birnstiel}, {Bai}, {Benisty}, {Hughes}, {{\"O}berg}, \&
  {Ricci}}]{2018ApJ...869L..41A}
{Andrews}, S.~M., {Huang}, J., {P{\'e}rez}, L.~M., {et~al.} 2018, \apj, 869,
  L41

\bibitem[{{Auffinger} \& {Laibe}(2018)}]{2018MNRAS.473..796A}
{Auffinger}, J., \& {Laibe}, G. 2018, \mnras, 473, 796

\bibitem[{{Bai} \& {Stone}(2010{\natexlab{a}})}]{2010ApJ...722.1437B}
{Bai}, X.-N., \& {Stone}, J.~M. 2010{\natexlab{a}}, \apj, 722, 1437

\bibitem[{{Bai} \& {Stone}(2010{\natexlab{b}})}]{2010ApJ...722L.220B}
---. 2010{\natexlab{b}}, \apjl, 722, L220

\bibitem[{{Birnstiel} {et~al.}(2010){Birnstiel}, {Dullemond}, \&
  {Brauer}}]{2010A&A...513A..79B}
{Birnstiel}, T., {Dullemond}, C.~P., \& {Brauer}, F. 2010, \aap, 513, A79

\bibitem[{{Birnstiel} {et~al.}(2012){Birnstiel}, {Klahr}, \&
  {Ercolano}}]{2012A&A...539A.148B}
{Birnstiel}, T., {Klahr}, H., \& {Ercolano}, B. 2012, \aap, 539, A148

\bibitem[{{Birnstiel} {et~al.}(2018){Birnstiel}, {Dullemond}, {Zhu}, {Andrews},
  {Bai}, {Wilner}, {Carpenter}, {Huang}, {Isella}, {Benisty}, {P{\'e}rez}, \&
  {Zhang}}]{2018ApJ...869L..45B}
{Birnstiel}, T., {Dullemond}, C.~P., {Zhu}, Z., {et~al.} 2018, \apj, 869, L45

\bibitem[{{Carrera} {et~al.}(2015){Carrera}, {Johansen}, \&
  {Davies}}]{2015A&A...579A..43C}
{Carrera}, D., {Johansen}, A., \& {Davies}, M.~B. 2015, \aap, 579, A43

\bibitem[{{Cazzoletti} {et~al.}(2018){Cazzoletti}, {van Dishoeck}, {Pinilla},
  {Tazzari}, {Facchini}, {van der Marel}, {Benisty}, {Garufi}, \&
  {P{\'e}rez}}]{2018A&A...619A.161C}
{Cazzoletti}, P., {van Dishoeck}, E.~F., {Pinilla}, P., {et~al.} 2018, \aap,
  619, A161

\bibitem[{{Dittrich} {et~al.}(2013){Dittrich}, {Klahr}, \&
  {Johansen}}]{2013ApJ...763..117D}
{Dittrich}, K., {Klahr}, H., \& {Johansen}, A. 2013, \apj, 763, 117

\bibitem[{{Draine}(2003)}]{2003ApJ...598.1026D}
{Draine}, B.~T. 2003, \apj, 598, 1026

\bibitem[{{Dr{\c a}{\.z}kowska} {et~al.}(2016){Dr{\c a}{\.z}kowska}, {Alibert},
  \& {Moore}}]{2016A&A...594A.105D}
{Dr{\c a}{\.z}kowska}, J., {Alibert}, Y., \& {Moore}, B. 2016, \aap, 594, A105

\bibitem[{{Dubrulle} {et~al.}(1995){Dubrulle}, {Morfill}, \&
  {Sterzik}}]{1995Icar..114..237D}
{Dubrulle}, B., {Morfill}, G., \& {Sterzik}, M. 1995, \icarus, 114, 237

\bibitem[{{Dullemond} {et~al.}(2018){Dullemond}, {Birnstiel}, {Huang},
  {Kurtovic}, {Andrews}, {Guzm{\'a}n}, {P{\'e}rez}, {Isella}, {Zhu}, {Benisty},
  {Wilner}, {Bai}, {Carpenter}, {Zhang}, \& {Ricci}}]{2018ApJ...869L..46D}
{Dullemond}, C.~P., {Birnstiel}, T., {Huang}, J., {et~al.} 2018, \apj, 869, L46

\bibitem[{{G{\'a}rate} {et~al.}(2019){G{\'a}rate}, {Birnstiel}, {Drazkowska},
  \& {Stammler}}]{2019arXiv190607708G}
{G{\'a}rate}, M., {Birnstiel}, T., {Drazkowska}, J., \& {Stammler}, S.~M. 2019,
  arXiv e-prints, arXiv:1906.07708

\bibitem[{{Henning} \& {Stognienko}(1996)}]{1996A&A...311..291H}
{Henning}, T., \& {Stognienko}, R. 1996, \aap, 311, 291

\bibitem[{{Huang} {et~al.}(2018{\natexlab{a}}){Huang}, {Andrews}, {Dullemond},
  {Isella}, {P{\'e}rez}, {Guzm{\'a}n}, {{\"O}berg}, {Zhu}, {Zhang}, {Bai},
  {Benisty}, {Birnstiel}, {Carpenter}, {Hughes}, {Ricci}, {Weaver}, \&
  {Wilner}}]{2018ApJ...869L..42H}
{Huang}, J., {Andrews}, S.~M., {Dullemond}, C.~P., {et~al.} 2018{\natexlab{a}},
  \apj, 869, L42

\bibitem[{{Huang} {et~al.}(2018{\natexlab{b}}){Huang}, {Andrews}, {P{\'e}rez},
  {Zhu}, {Dullemond}, {Isella}, {Benisty}, {Bai}, {Birnstiel}, {Carpenter},
  {Guzm{\'a}n}, {Hughes}, {{\"O}berg}, {Ricci}, {Wilner}, \&
  {Zhang}}]{2018ApJ...869L..43H}
{Huang}, J., {Andrews}, S.~M., {P{\'e}rez}, L.~M., {et~al.} 2018{\natexlab{b}},
  \apj, 869, L43

\bibitem[{{Isella} {et~al.}(2016){Isella}, {Guidi}, {Testi}, {Liu}, {Li}, {Li},
  {Weaver}, {Boehler}, {Carperter}, {De Gregorio-Monsalvo}, {Manara}, {Natta},
  {P{\'e}rez}, {Ricci}, {Sargent}, {Tazzari}, \&
  {Turner}}]{2016PhRvL.117y1101I}
{Isella}, A., {Guidi}, G., {Testi}, L., {et~al.} 2016, \prl, 117, 251101

\bibitem[{{Ivezic} {et~al.}(1997){Ivezic}, {Groenewegen}, {Men'shchikov}, \&
  {Szczerba}}]{1997MNRAS.291..121I}
{Ivezic}, Z., {Groenewegen}, M.~A.~T., {Men'shchikov}, A., \& {Szczerba}, R.
  1997, \mnras, 291, 121

\bibitem[{{Johansen} {et~al.}(2007){Johansen}, {Oishi}, {Mac Low}, {Klahr},
  {Henning}, \& {Youdin}}]{2007Natur.448.1022J}
{Johansen}, A., {Oishi}, J.~S., {Mac Low}, M.-M., {et~al.} 2007, \nat, 448,
  1022

\bibitem[{{Johansen} \& {Youdin}(2007)}]{2007ApJ...662..627J}
{Johansen}, A., \& {Youdin}, A. 2007, \apj, 662, 627

\bibitem[{{Johansen} {et~al.}(2009){Johansen}, {Youdin}, \& {Mac
  Low}}]{2009ApJ...704L..75J}
{Johansen}, A., {Youdin}, A., \& {Mac Low}, M.-M. 2009, \apjl, 704, L75

\bibitem[{{Kataoka} {et~al.}(2015){Kataoka}, {Muto}, {Momose}, {Tsukagoshi},
  {Fukagawa}, {Shibai}, {Hanawa}, {Murakawa}, \&
  {Dullemond}}]{2015ApJ...809...78K}
{Kataoka}, A., {Muto}, T., {Momose}, M., {et~al.} 2015, \apj, 809, 78

\bibitem[{{Liu}(2019)}]{2019arXiv190400333L}
{Liu}, H.~B. 2019, arXiv e-prints, arXiv:1904.00333

\bibitem[{{Lynden-Bell} \& {Pringle}(1974)}]{1974MNRAS.168..603L}
{Lynden-Bell}, D., \& {Pringle}, J.~E. 1974, \mnras, 168, 603

\bibitem[{{Macias} {et~al.}(2019){Macias}, {Espaillat}, {Osorio}, {Anglada},
  {Torrelles}, {Carrasco-Gonzalez}, {Flock}, {Linz}, {Bertrang}, {Henning},
  {Gomez}, {Calvet}, \& {Dent}}]{2019arXiv190707277M}
{Macias}, E., {Espaillat}, C., {Osorio}, M., {et~al.} 2019, arXiv e-prints,
  arXiv:1907.07277

\bibitem[{{Mathis} {et~al.}(1977){Mathis}, {Rumpl}, \&
  {Nordsieck}}]{1977ApJ...217..425M}
{Mathis}, J.~S., {Rumpl}, W., \& {Nordsieck}, K.~H. 1977, \apj, 217, 425

\bibitem[{{Ormel}(2017)}]{2017ASSL..445..197O}
{Ormel}, C.~W. 2017, in Astrophysics and Space Science Library, ed. M.~{Pessah}
  \& O.~{Gressel}, Vol. 445, 197

\bibitem[{{Pinilla} {et~al.}(2012){Pinilla}, {Benisty}, \&
  {Birnstiel}}]{2012A&A...545A..81P}
{Pinilla}, P., {Benisty}, M., \& {Birnstiel}, T. 2012, \aap, 545, A81

\bibitem[{{Pinte} {et~al.}(2018){Pinte}, {Price}, {M{\'e}nard}, {Duch{\^e}ne},
  {Dent}, {Hill}, {de Gregorio-Monsalvo}, {Hales}, \&
  {Mentiplay}}]{2018ApJ...860L..13P}
{Pinte}, C., {Price}, D.~J., {M{\'e}nard}, F., {et~al.} 2018, \apj, 860, L13

\bibitem[{{Ricci} {et~al.}(2018){Ricci}, {Isella}, {Andrews}, {Birnstiel},
  {Cuzzi}, {D'Angelo}, {Dong}, {Dutrey}, {Ercolano}, {Estrada}, {Flock}, {Li},
  {Liu}, {Lyra}, {Oberg}, {Okuzumi}, {Perez}, {Turner}, {van der Marel},
  {Wilner}, {Youdin}, \& {Zhu}}]{2018arXiv180304467R}
{Ricci}, L., {Isella}, A., {Andrews}, S.~M., {et~al.} 2018, arXiv e-prints,
  arXiv:1803.04467

\bibitem[{Rice {et~al.}(2006)Rice, Armitage, Wood, \&
  Lodato}]{2006MNRAS.373.1619R}
Rice, W. K.~M., Armitage, P.~J., Wood, K., \& Lodato, G. 2006, MNRAS, 373, 1619

\bibitem[{{Schoonenberg} \& {Ormel}(2017)}]{2017A&A...602A..21S}
{Schoonenberg}, D., \& {Ormel}, C.~W. 2017, \aap, 602, A21

\bibitem[{{Schoonenberg} {et~al.}(2018){Schoonenberg}, {Ormel}, \&
  {Krijt}}]{2018A&A...620A.134S}
{Schoonenberg}, D., {Ormel}, C.~W., \& {Krijt}, S. 2018, \aap, 620, A134

\bibitem[{{Schreiber} \& {Klahr}(2018)}]{2018ApJ...861...47S}
{Schreiber}, A., \& {Klahr}, H. 2018, \apj, 861, 47

\bibitem[{{Sekiya} \& {Onishi}(2018)}]{2018ApJ...860..140S}
{Sekiya}, M., \& {Onishi}, I.~K. 2018, \apj, 860, 140

\bibitem[{{Shakura} \& {Sunyaev}(1973)}]{1973A&A....24..337S}
{Shakura}, N.~I., \& {Sunyaev}, R.~A. 1973, \aap, 500, 33

\bibitem[{{Taki} {et~al.}(2016){Taki}, {Fujimoto}, \&
  {Ida}}]{2016A&A...591A..86T}
{Taki}, T., {Fujimoto}, M., \& {Ida}, S. 2016, \aap, 591, A86

\bibitem[{{Teague} {et~al.}(2018){Teague}, {Bae}, {Bergin}, {Birnstiel}, \&
  {Foreman-Mackey}}]{2018ApJ...860L..12T}
{Teague}, R., {Bae}, J., {Bergin}, E.~A., {Birnstiel}, T., \& {Foreman-Mackey},
  D. 2018, \apj, 860, L12

\bibitem[{{Tsukagoshi} {et~al.}(2016){Tsukagoshi}, {Nomura}, {Muto}, {Kawabe},
  {Ishimoto}, {Kanagawa}, {Okuzumi}, {Ida}, {Walsh}, \&
  {Millar}}]{2016ApJ...829L..35T}
{Tsukagoshi}, T., {Nomura}, H., {Muto}, T., {et~al.} 2016, \apj, 829, L35

\bibitem[{{Warren} \& {Brandt}(2008)}]{2008JGRD..11314220W}
{Warren}, S.~G., \& {Brandt}, R.~E. 2008, Journal of Geophysical Research
  (Atmospheres), 113, D14220

\bibitem[{{Windmark} {et~al.}(2012){Windmark}, {Birnstiel}, {G{\"u}ttler},
  {Blum}, {Dullemond}, \& {Henning}}]{2012A&A...540A..73W}
{Windmark}, F., {Birnstiel}, T., {G{\"u}ttler}, C., {et~al.} 2012, \aap, 540,
  A73

\bibitem[{{Yang} {et~al.}(2017){Yang}, {Johansen}, \&
  {Carrera}}]{2017A&A...606A..80Y}
{Yang}, C.~C., {Johansen}, A., \& {Carrera}, D. 2017, \aap, 606, A80

\bibitem[{{Yang} {et~al.}(2018){Yang}, {Mac Low}, \&
  {Johansen}}]{2018ApJ...868...27Y}
{Yang}, C.-C., {Mac Low}, M.-M., \& {Johansen}, A. 2018, \apj, 868, 27

\bibitem[{{Youdin} \& {Johansen}(2007)}]{2007ApJ...662..613Y}
{Youdin}, A., \& {Johansen}, A. 2007, \apj, 662, 613

\bibitem[{{Youdin} \& {Goodman}(2005)}]{2005ApJ...620..459Y}
{Youdin}, A.~N., \& {Goodman}, J. 2005, \apj, 620, 459

\bibitem[{{Youdin} \& {Lithwick}(2007)}]{2007Icar..192..588Y}
{Youdin}, A.~N., \& {Lithwick}, Y. 2007, \icarus, 192, 588

\bibitem[{{Zhu} {et~al.}(2019){Zhu}, {Zhang}, {Jiang}, {Kataoka}, {Birnstiel},
  {Dullemond}, {Andrews}, {Huang}, {Perez}, {Carpenter}, {Bai}, \&
  {Wilner}}]{2019arXiv190402127Z}
{Zhu}, Z., {Zhang}, S., {Jiang}, Y.-F., {et~al.} 2019, arXiv e-prints,
  arXiv:1904.02127

\end{thebibliography}



\end{document}